# Photoplastic effects in chalcogenide glasses: A review

**S. N. Yannopoulos*,[1] and M. L. Trunov[2]**

[1] Foundation for Research and Technology Hellas - Institute of Chemical Engineering and High Temperature Chemical Processes, (FORTH/ICE-HT), P.O. Box 1414, GR-26504, Patras, Greece
[2] Uzhgorod National University, ul. Pidhirna 46, Uzhgorod, 88000 Ukraine

* Corresponding author: e-mail sny@iceht.forth.gr, Phone: +30 2610 965252, Fax: +30 2610 965223

A synopsis of the various photoinduced changes of rheological, mechanical and elastic properties is presented in the first part of the article. After a critical appraisal of a large body of experimental data it suggested that the photoviscous effect, that is, the athermal decrease of viscosity of a non-crystalline chalcogenide upon illumination is the key for a plethora of photoinduced effects reported so far in the literature under different names. Morphic effects (shape or surface morphology) may appear either in the presence or absence of external mechanical stimuli leading to the fabrication of a variety of technologically important photoprocessed structures. A few representative examples of photoplastic effects are described, in the second part of the paper, in some detail based on information provided by *in situ* Raman scattering and nanoindentation experiments.

**1. Introduction** More than fifty years ago Goryunova and Kolomiets reported the discovery of the semiconducting properties of vitreous chalcogenides or chalcogenide glasses (ChGs) [1, 2]. However, this finding did not go unchallenged. Eminent solid state physicists [3] raised serious objections, which were soon diminished, supporting that without the long range order (lattice) a three dimensional system would not preserve a bandgap. At approximately the same time S. R. Ovshinsky [4] independently set-off a systematic investigation of the electrical properties of disordered chalcogenides. His main purpose was to study switching effects. His discoveries and the noting of new effects led to the establishment of the so-called Ovonic research.

Over the years, research on ChGs grew enormously due to the strong interest in these materials from both the scientific and technological viewpoint [4-8]. Applications of ChGs have mainly been based on their transparency to infrared light (passive) and their ability to “respond” to near-bandgap illumination (active). The latter, i.e. the external-stimuli control of the structure and properties of ChGs is perhaps scientifically the most interesting case in view of the challenges it poses for a microscopic understanding of photoinduced phenomena [8].

Light is a convenient stimulus that has been extensively employed to induce, and to detect, subtle structural modifications in amorphous media [6,7]. Such photoinduced structural changes are ultimately responsible for the changes in many macroscopically observed properties. Photoinduced structural changes are classified as metastable phenomena because they last for time intervals exceeding the observation time, in contrast to transient effects occurring on much shorter time scale [6,8]. Non-radiative recombination of the photo-excited carrier is the key process that leads to photoinduced structural changes and subsequently to modifications of other properties. Several physical and chemical properties of ChGs, such as mechanical/rheological, optical, electrical, thermodynamic, etc., experience changes after illumination.

Several photoinduced phenomena have been thoroughly studied and sufficiently understood. However, no clear picture has yet emerged for photoinduced changes in mechanical/rheological properties for two main reasons. First, such studies have received considerably less attention compared to those concerned with optical or electrical properties. Second, and more important, there has been in the literature confusion about the terminology and nature various photoinduced effects concerning mechanical/rheological properties. As will become clear in the next section, modifications – in relation to dark properties – of several properties such as viscosity, glass transition temperature ($T_g$), elastic constants, microhardness, stress re-laxation, surface morphology etc., have been studied under illumination conditions. Perhaps unduly, each of the above cases was considered as a new effect and hence several terms have been introduced and used, such as photodeformation, photoviscous, photoplastic, photohardening, photofluidity, photoinduced structure (stress) relaxation, photoinduced mass transport, photoinduced relief grating, photomelting, optomechanical effect, etc.

A critical appraisal of the abovementioned effects points to a common origin, i.e. the photoinduced transformation of the material from a solid glassy state to a highly viscous "fluid" which practically induces a transition form a brittle state to a plastic one. This is the key that renders the material easily amenable to change its shape and/or "size" under proper conditions such as external forces, prolonged illumination etc. As long as only illumination is involved the term "photoviscous" would be an appropriate one since it pertains to the origin of the effect. In the presence of an external mechanical stimulus the term "photoplastic" seems more suitable because morphic or shape effects emerge making the effect obvious. However, morphic effects can also appear solely under the action of light (surface relief deformations) which makes the above distinction not clear.

It is worth noticing that most of the effects mentioned above and analyzed in more detail in Section 2 are not unique to amorphous chalcogenides. Corresponding effects have been observed and studied in detail in organic light-sensitive materials. Polymers either in the glassy or the liquid crystalline state, which contain azobenzene molecules (chromophores) exhibit a dazzling variety of photoinduced vectoral (light polarization dependent) effects owing to *trans↔cis* isomerization which leads to a large change in the molecular length [9(a)]. Cycling of the isomerization transition leads through conformational changes of the azobenzene molecules to molecular motion at the nanoscale. The cumulative effect of such motions lead to spectacular *fast* photomechanical effects in large area pieces (1mm × 1mm × 10μm) such as anisotropic folding and unfolding [9(b), (c)] as well as anisotropic photoinduced mass transport and photohealing effects [9(d)] in channels inscribed on film's surface under illumination with linearly polarized light. A comparison of photoinduced effects between chalcogenides and azobenzene-containing materials has been reported elsewhere [9(e)]. It is remarkable the similarity in terminology used to describe photoinduced effects in chalcogenide and azobenzene materials but these two fields of research seem to proceed in parallel without considerable interaction.

## 2 A brief history of athermal photoviscous/photoplastic effects

**2.1 The rise of photoplastic effects** It is remarkable that ninety years ago O. U. Vonwiller published an abstract [10] on his studies of the elastic properties of vitreous Se. He observed a continuous yielding of the material under the action of distorting forces. The yielding increased appreciably when Se was illuminated. This photoinduced effect on viscosity was also observed, albeit severely suppressed in magnitude, in the crystalline state of elemental Se [10]. Not many details about this effect were given.

**2.2 Photoplastic effects of crystalline semiconductors** Athermal light-induced plasticity in crystalline semiconductor is known since 1957 [11(a)] when it was reported that *in situ* light illumination causes strong decrease of hardness over a small depth of the crystal surface. This phenomenon, which was called *photomechanical effect*, takes place for infrared and ultraviolet light for InAs and InSb crystals. It was originally attributed to the unblocking and increased mobility of dislocations due to photoinduced increase of carrier concentration. The effect was later demonstrated for several single crystals [11(b)]. The photomechanical effect was attributed to a mechanism according to which microhardness decreases owing to chemical bond weakening and isotropization of the crystal caused by the "photogenerated antibonding quasiparticles" [11(b)].

The opposite effect, the athermal photohardening of a semiconductor crystals has also intensively studied after the pioneering work of Osipyan and coworkers [12] who demonstrated that the elastic properties of crystalline CdS exhibits strengthening upon illumination. A strong reversible increase in flow stress during illumination with light of proper wavelength was observed. The term *photoplastic effect* was introduced. It was shown that the effect depends on the illumination power, temperature, and light wavelength. With increasing illumination power the effect saturates while the plastic deformation exhibits anomalous behavior since it decreases considerably with increasing temperature. Studies of the spectral dependence revealed that the effect is maximized when the illumination wavelength is close to the bandgap of the crystal. The effect was observed in several other crystals including ZnO [13(a),(b)] and Si [13(c)]. Several explanations were offered to account for this photohardening effect including the change of the conditions for the motion of dislocations during illumination due to the change of free electrons interacting with the moving dislocations [12]. Bandgap illumination generates electron-hole pairs; holes are captured by traps and produce doubly charged ions. The latter interact more strongly with dislocations thus raising flow stress by dislocation locking [13(a), (b)].

**2.3 Photodeformation studies of chalcogenide films** Photoinduced structure relaxation was demonstrated in chalcogenide amorphous films such as $As_{20}S_{80}$ and $Sb_2S_3$ deposited on mica substrates [14]. The effect became evident through the reversible deformation (bending) of the chalcogenide film upon band gap light illumination. The effect was ascribed to the thermal expansion of the film due to absorption of the exciting light. However, the authors have stressed that the temperature increase due to bandgap illumination may have a non-negligible effect on photoinduced structure relaxation. Similar effects were also observed in several chalcogenide films ($As_2S_3$, $As_2Se_3$, $As_{40}Se_{50}Ge_{10}$, $As_{40}S_{25}Se_{25}Ge_{10}$, $As_{40}S_{10}Se_{40}Ge_{10}$) by Igo et al. [15(a)]. A more detailed study on photoinduced stress relaxation for a-Se was presented by Koseki and Odajima [15(b)] who measured the deformation of a-Se/mica films bilayer. The photoinduced stress relaxation was attributed to non-radiative recombination of photo-excited electron-hole pairs in the framework of the configurational coordinate diagram. A more sophisticated apparatus for studying photodeformation in amorphous films and bulk ChGs – Se and $As_2S_3$ – has been reported by Rykov and coworkers [16]. Valence alternations pairs were invoked to account for the observed effects. Exploiting the different response of an a-$As_{50}Se_{50}$ film (deposited on a cantilever's surface) under illumination of polarized light it was shown that reversible contractions and dilations of the film/cantilever system take place depending on the polarization direction of the incident light [17].

**2.4 Photoinduced effects on elastic constants and $T_g$** Photoinduced increase in ultrasound velocities of acoustic surface waves and hence of the elastic constants have been reported for as-deposited $As_2S_3$ films under bandgap illumination [18(a)]. More recent studies have been conducted in Ge-Se glasses using near-bandgap Brillouin scattering [18(b)]. A significant light-induced softening of the longitudinal elastic constant was observed over a narrow compositional range close to the mean-field rigidity transition composition.

Detailed studied of the photo-dependence of sub-$T_g$ relaxations in a-Se films have been conducted by Larmagnac et al. [19]. Studying the ageing of thin films of a-Se under illumination it was found that light produces a significant increase of the structural relaxation time below $T_g$. In particular, $T_g$ was found to increase with ageing time when a-Se was illuminated with above-bandgap light at various energies. The higher the light energy the larger increase of the $T_g$. Given the small penetration depth at such energies, the fact that only a very small part of the film volume was illuminated led the authors [19] to conclude that the photo-excited carriers diffuse over the total volume of the sample which is measured by thermal analysis. Interestingly, Koseki and Odajima [20] reported that the thermal history of a-Se is crucial for the positive or negative change of $T_g$ after illumination. In particular, the $T_g$ of the illuminated material increased in comparison with the freshly prepared not-illuminated film, while the converse is true, i.e. illumination of a film aged at ambient temperature for one week causes decrease of the $T_g$.

**2.5 Photoplastic effects: in situ and ex situ indentation studies** Several studies of photoinduced changes in mechanical properties of chalcogenide films were conducted by means of monitoring microhardness changes after illumination. Early studies on $As_{60}Se_{40}$ [21] revealed that illuminated regions of the films were less amenable to be scratched than not-illuminated regions giving the sense of hardening of the post-illuminated material. No quantitative measurements of microhardness were performed [21]. It was shown that the effect is limited within a thin layer of the film (due to above bandgap illumination) and annealing near $T_g$ erases photohardening. To account for the effect the authors [21] invoked the mechanism of photodecomposition as sketched by Berkes *et al.* [22].

The first *in situ* study of illumination and microhardness determination on thick films of a-Se and bulk glassy g-$As_2Se_3$ was reported by Deryagin *et al.* [23]. They found a significant decrease of microhardness upon illumination which amounts to ~20-25% for a-Se and ~50% for g-$As_2Se_3$. Given this finding, and being influenced by the corresponding result of the photomechanical effect in crystals reported in [11(a)], Deryagin *et al.* used the same term to describe their findings. The authors found negative evidence for a spectral dependence of the magnitude of the photoinduced changes in microhardness. To account for their finding, which is actually the first quantitative study of its kind, intramolecular bond changes were invoked. More detailed quantitative studies (Vickers method) on the photohardening effect were performed by Kolomiets *et al.* [24] for $As_{60}Se_{40}$ films. Microhardness was found to oscillate reversibly between two well-defined values upon repeated illumination and annealing.

Subsequent studies for binary $As_xS_{100-x}$ ($15 \le x \le 50$) and $As_zSe_{100-z}$ ($40 \le z \le 80$) glasses showed that the change in Vickers microhardness in films (irradiated up to the level of saturation of the optical transmission changes) exhibits maxima near x=45 and z=55 [25]. These observations were accounted for by photoinduced polymerization of As-Se molecular units (at x=55) by means of the appearance of valence alternation pairs of under- and over-coordinated As and Se atoms, respectively. Anomalous temperature dependence of the microhardness of irradiated films was observed over a temperature range up to $T_g$ [25].

A systematic study of photoplastic changes using *in situ* microindentation techniques has been undertaken by Trunov and coworkers [26]. The kinetics of photoinduced structural relaxation in amorphous As-S films was investigated by monitoring the changes that occur in their intrinsic and extrinsic stresses during the process of illuminating with bandgap light. Photoinduced stress relaxation was observed under illumination where the magnitude of the stress decreases to zero. A macroscopic model of this phenomenon was based on the photoinduced decrease of viscosity to $10^{12}$-$10^{13}$ Poise. The viscosity reduction is achieved by a purely optical (athermal) way and the magnitude is very close to the magnitude of viscosity near $T_g$.

In the aforementioned *ex situ* studies (where the effect of illumination on hardness was evaluated after illumination) a photohardening of glass structure was observed. The formation of additional intermolecular bonds after cessation of the irradiation, which could transform the photo-processed structure to more rigid configuration, seems the most probable candidate for the microscopic origin of the effect. On the contrary, *in situ* studies (measurements of mechanical properties during illumination) revealed a photosoftening of glass structure due to photoinduced increase of fluidity. The origin of this effect is described in the next section.

**2.6 Photoviscous effects in bulk ChGs** A very interesting series of experiments carried out by Nemilov and Tagantsev dealt with the *in situ* observation of viscosity changes of several ChGs under illumination [27]. The term *photoviscous effect* was coined to account for the athermal decrease of the material's equilibrium viscosity under the action of light. In a very elegant way the authors managed to separate genuine photoinduced changes on viscosity from those arising from temperature nonuniformity due to illumination heating. From Nemilov's work, unjustifiably overlooked in several subsequent studies, information was gained about the dependence of the photoviscous effect on: (i) the intensity of the incident light, (ii) the wavelength (covering a spectral dependence from 300 to 1200 nm, i.e. from below to above bandgap conditions) of the incident light, and (iii) the glass transition temperature. The $T_g$ was found to drastically decrease upon illumination. The photoviscous effect was found to depend exponentially on light intensity. An interesting finding was that the light influence on viscosity decreases with increasing temperature, signaling anomalous temperature dependence. This result was attributed to the decrease of the activation free energy of viscous flow upon illumination.

The spectral sensitivity of the photoviscous effect revealed an asymmetric maximum near the energy of the bandgap of the glass. No photoviscous effect is observed in the region of transparency and in the region of deep-band transitions. It appears here rather peculiar why the photoviscous effect in bulk glasses (1 mm thick) is facilitated for illumination energies near the bandgap where the penetration depth of the incident light is confined to the first few μm of the material. However, the fact that the photoviscous effect is a bulk phenomenon implies that the photoinduced changes occurring in a thin surface layer spread into the depth of the material and facilitate relaxation of the stresses (viscosity decrease). Therefore, the process seems to be dominated by the self-diffusion of photoinduced structural defects. This suggestion is in line with the idea of diffusion of photoinduced defects in a-Se [19] at ambient temperature, as described in Section 2.4. The high temperature at which the experiments by Nemilov and Tagantsev [27] were conducted implies a facilitation of the diffusion mechanism. Theoretical concepts on the mechanism of viscous flow that had earlier developed by Nemilov were employed to describe the photoviscous effect [27(c)]. In particular, it was considered that the part of the activation energy of viscous flow related to the barrier of configurational changes becomes lower under illumination thus facilitating flow events. Finally, similar conclusions as those presented above were reached in a recent study of the photoviscous effect in bulk glassy g-Se [28].

**2.7 Photoinduced mass transport: relief gratings and healing effects** Kolomiets was presumably the first who reported photoinduced mass transport studying the kinetics of "healing" processes of scratches inscribed on the surface of chalcogenide films [29]. The temperature at which a channel scratched on a chalcogenide film fills-up due to material "flow" was found to change in a reversible way during illumination and annealing cycles. The formation of giant relief modulations in a-$As_2S_3$ has been reported upon intense bandgap light illumination [30(a)]. This *optically field-induced mass transport effect* was attributed to the photoinduced softening of the glass matrix (recall photoviscous effect) with the concomitant formation of defects with enhanced polarizability. These defects drift under the optical field gradient force forming surface relief modulations. The effect is polarization dependent and is realized by both single beam and two cross-polarized beam experiments. It was considered that mass transport takes place due to the electric field gradient force of the excitation light exerted on the photogenerated defects with high dielectric polarizabilities. A similar effect was reported for a-Se, i.e. emergence of relief deformations under bandgap illumination at low temperature [30(b)]. However, the observations were assigned to *photomelting* proceeding via an interchain bond breaking mechanism. It seems plausible that the photoviscous effect accompanied with mass transport might be related to the photomelting effect.

Photoinduced relief deformations are very common in azobenzene containing materials which occur due to the light-activated interconversion between the two geometrical isomers. The *trans*→*cis* isomerization takes place under irradiation and the back-conversion can occur either by irradiation or thermally [30(c)].

Other systematic studies have demonstrated that giant, laser induced, polarization-sensitive mass-transport occurs under irradiation at moderate light intensity and long exposure times (20-360 mW $cm^{-2}$ and 400-600 min) [31, 32]. Anisotropic relief deformations and mass transport of the material were induced in As-S(Se) films by illumination of unfocused linearly polarized light. The magnitude of the effect was monitored through the kinetics (time dependence) of expansion of a channel inscribed on a chalcogenide film's surface under uniform illumination. Details about photoplastic phenomena described here will be analyzed in more detail in Section 5.

**2.8 Recent studies of morphic effects in bulk ChGs** Exploiting the photoviscous effect discovered my Nemilov and coworkers [27], i.e. the reduced viscosity in the course of illumination, the effect of permanent deformation (shape change) of a-$As_2S_3$ (in the form of fibers and flakes) under the combined action of illumination and mechanical stress was reported [33(a)]. The effect was termed as *photoinduced fluidity* although the glass does not really flow; instead it deforms (fiber elongation or flake bending) under the action of an external stress, justifying the term photo-induced ductility as a more proper one. The effect exhibits anomalous temperature dependence [33(a)], in the case of flakes, as was previously reported for the photo-hardening [25] and the photoviscous effect [27] pointing to a common microscopic origin of these effects. However, qualitative and quantitative differences of the temperature dependence of the effect were found between flakes and fibers [33(b)]. The temperature dependence of the photoviscous effect in fibers is not anomalous, thus contradicting the results on flakes [33(b)]. The glass preparation conditions (evaporation for flakes, drawing from melt for fibers) must play some important role for the observed difference because vacuum evaporation leads to more inhomogeneous (nanophase separated) structures. On the other hand, fibers are homogeneous and, unless they are prepared at extreme extrusion rates, are devoid of oriented substructures (see Section 3.1 for details).

Fritzsche [34(a)] provided an explanation for the microscopic origin of the effect employing the self-trapped exciton idea. In that sense, the photo-excited electron-hole pairs are considered to recombine non-radiatively through an intermediate transient exciton state. The final recombination may yield a bonding arrangement different from the initial configuration. Then the observed macroscopic changes in the fluidity arise from the cumulative effect of the local configuration changes in line with Nemilov's approach [27]. One may also recall here the "slip and repulsion" model that was proposed to account for the photoinduced effect such volume expansion [34(b)]. In the context of this model, the volume expansion is the result of the electrostatic repulsion forces between adjacent layers after excitation. These forces built up due to the lower mobility of electrons compared to holes. After volume expansion the mutual slipping between adjacent of the layers is facilitated. However, subsequent calculations [34(c)] have shown that the Coulomb interaction between the photogenerated carriers is not capable of explaining the experimentally observed magnitude of deformation of illuminated chalcogenides.

A systematic experimental *in situ* study of the microscopic origin of photoviscous effects was undertaken using Raman scattering [35-41]. The structural mechanism underling photoviscous effects was elucidated by tracking changes in the vibrational modes of the glass, using Raman scattering, during the time that the effect takes place. Raman spectroscopy is an ideally suitable probe that can provide information concerning both short and intermediate length scales. A selection of these results is briefly summarized in the next Section.

## 3 Photoplastic effects in As-S glasses: Raman scattering studies on uniaxially stressed fibers

### 3.1 Orientational order enhancement in $As_2S_3$ glass structure as an indicator of photoplastic effects

Fibers were more preferable than flakes in performing Raman experiments for two main reasons. First, fibers have a well-defined cylindrical shape which makes possible the accurate determination of their diameters and hence the correct calculation of the applied stress. Second, fiber's structure is a better representation of bulk glass structure in comparison with the evaporated thick films. One could argue that the fiber drawing procedure from the melt might induce some ordering in the glass structure as has been reported for $As_2Se_3$ [42]. In that work, fibers were pulled from the melt at a speed greater than 100 m/min with a final fiber diameter d ≈ 1 μm, while in our experiment much lower drawing rates are used (at least one order of magnitude) leading to a fiber diameter d ≈ 100-300 μm. More importantly, the drawing-induced "crystallites" were shown to survive only at 77 K, and seemed to anneal to the amorphous phase at 300 K [42].

Several sets of Stokes–side, polarized (VV: vertical electric filed of the incident radiation - vertical analysis of the electric filed of the scattered radiation) and depolarized (HV: horizontal electric filed of the incident radiation - vertical analysis of the electric filed of the scattered radiation) Raman spectra from fibers of various diameters have been recorded as a function of uniaxial stress [35-41]. The fiber axis was always fixed at a perpendicular (V) configuration relative to the scattering plane. It was shown that the photoplastic effect in fibers was induced by a range of laser power densities on the scattering volume. In our studies we have used a power density ~ 20 W $cm^{-2}$. This value is close to the lowest threshold above which the effect was measurable. In addition, it was chosen because it resulted in a slow kinetics of the effect in comparison with the time needed to record a set of VV and VH spectra. Therefore, no appreciable change took place during this recording time of the Raman spectra at constant stress magnitude.

The most obvious feature made out from Raman spectra is the monotonic increase of the depolarization ratio $\rho = I^{HV} / I^{VV}$ when increasing the magnitude of the external stress, *S*. The spectral region 260–430 $cm^{-1}$ (which includes symmetric and antisymmetric stretching vibrational modes of $AsS_3$ pyramidal units) is a characteristic one where the above observation comes from. This specific dependence of $\rho$ on *S* was observed for many fibers of various diameters as illustrated in Fig. 1 where values of $\rho$ represent the integrated area of the 260–430 $cm^{-1}$ spectral region (the data of Fig. 1 of Sulfur-rich glasses will be discussed below). This finding shows that the measured effect is well reproduced and repeatable justifying the use of $\rho$ as a reliable indicator apt to quantify the structural changes that occur in the glass structure during the illumination/stretching procedure.

It is important to note that the roles of the two stimuli, i.e. irradiation and stress on the changes of $\rho$ were examined independently. First, it was found that $\rho$ remains constant, regardless of the exposure time, if no stress is applied as shown in Fig. 2. Second, no plastic changes were observed in experiments were fibers were stressed in dark and the Raman spectra were recorded at selected time intervals while ceasing stressing. Thus, the changes of $\rho$ shown in Fig. 1 arise from the *combined* effect of illumination and stress application. This implies that illumination alone does not engender flow; it rather brings the structure to a state amenable to flow (photoviscous effect). The glass plastically deforms (photoplastic effect) with the aid of the external mechanical stimulus. Obviously, the term photoinduced fluidity is unjustifiable and might be misleading for this effect.

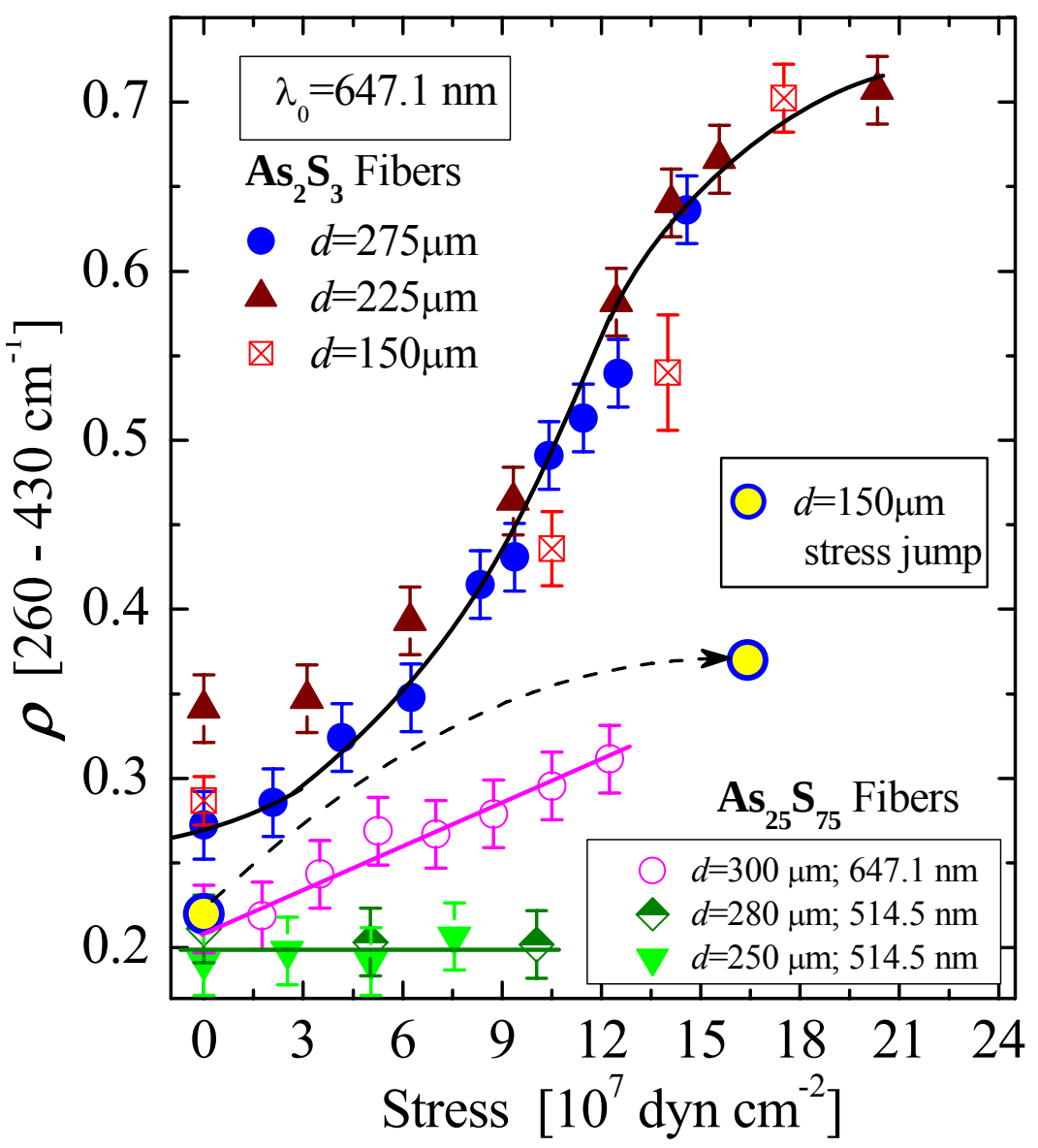

**Figure 1** Stress dependence of the depolarization ratio, a parameter that quantifies photoplastic effects, for $As_2S_3$ and $As_{25}S_{75}$ fibers of various diameters. The results are obtained at ambient temperature for different illumination sources including sub-bandgap and near-bandgap effects [36, 40].

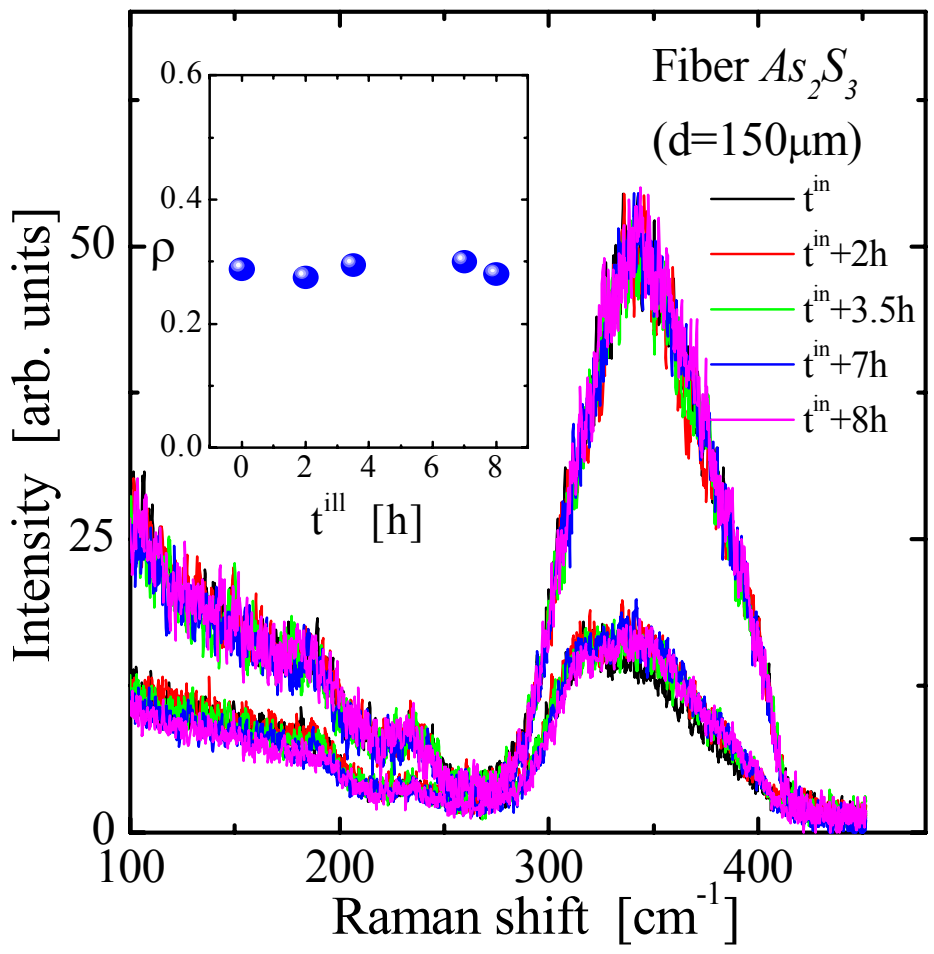

**Figure 2** Representative VV and HV Stokes-side Raman spectra of an $As_2S_3$ fiber as a function of time in the absence of uniaxial stress. The spectra have been normalized to the VV intensity peak maximum at ~350 $cm^{-1}$. Inset: Dependence of $\rho$ on illumination time.

Figure 1 reveals an interesting effect not reported in our previous studies [35-41]. A stress-jump experiment was performed where an initially unstressed fiber was subjected to a high value of $S$ close to that where saturation effects in $\rho$ were observed. The increase of $\rho$ is moderate in the stress-jump experiment revealing that the stress-history is important. Because the time-dependence of the effect at the low power density used for the experiment is not important it seems that there is a hierarchy of structural mechanisms involved that become activated during the gradual application of the uniaxial stress.

**3.2 Temperature dependence of the photoplastic effect** The temperature dependence of photoplastic effects has been frequently investigated by several authors [25, 27, 33] since it can provide support for the athermal nature of the effect. In all the above cases anomalous behavior is exhibited against temperature variations, namely the effect becomes less appreciable at higher temperatures. These findings provoked the study of vibrational modes at high temperatures [38, 39] undertaking a Raman spectroscopic study of $As_2S_3$ fibers over the temperature interval 20 – 120 °C. We have thus extended the information over a temperature range much broader than that used elsewhere [33(a)], however, not approaching $T_g$ in order to avoid heat-induced plasticity. Analyzing polarized and depolarized Raman spectra at various temperatures it is revealed that the rate of increase of the depolarization ratio is not constant, even non–monotonic, as a function of the temperature, as Fig. 3 illustrates. We observe that at 40 °C the difference between the magnitudes of the depolarization ratio at maximum stress relative to that in the absence of stress is of about 50% of the corresponding difference of the room temperature curve. Subsequent temperature rise, up to 60 °C, renders the effect even less appreciable. Evidently, photoplastic effects seem to be hindered at higher temperature as pointed our in previous studies [25, 27, 33(a)].

This picture changes drastically when the measurements are extended over a wider temperature range than that of Ref. [33(a), (b)]. Indeed, as it is illustrated in Fig. 3 the magnitude of $\rho$ increases again at 90 °C, where the $\rho$ vs. $S$ curve essentially coincides with that of the room temperature study. A further increase of temperature to 120 °C leads to more noticeable changes of the depolarization ratio even at low values of the applied stress. Bearing in mind that the glass transition temperature of the material ($T_g \approx 210$ °C) is still well above the highest employed temperature in our experiment, it seems highly unlikely that a substantial softening of the glass structure at 120 °C takes place.

To preclude any possibility of a "direct" role or thermal structure softening at the highest temperature reached in the present work we undertook the following experiment. A fiber was stressed ($8\times10^7$ dyn $cm^{-1}$) with the laser "off" at 120 °C for 3 h. Then, the fiber was cooled down to room temperature and the magnitude of the depolarization ratio was found practically similar to that measured before the thermal treatment, see Fig. 4. This asserts that at least up to 120 °C no thermal softening occurs and thus the features illustrated in Fig. 3 are purely photoinduced. Finally, it would be useful to note here that the narrowing of the badgap caused by the temperature rise up to 120 °C cannot affect the observed behavior of the depolarization ratio. Indeed, the band-gap energy change is of the order of 3% (see [39] for details) which is too small a difference to account for the immense, even non-monotonic, changes of the depolarization ratio.

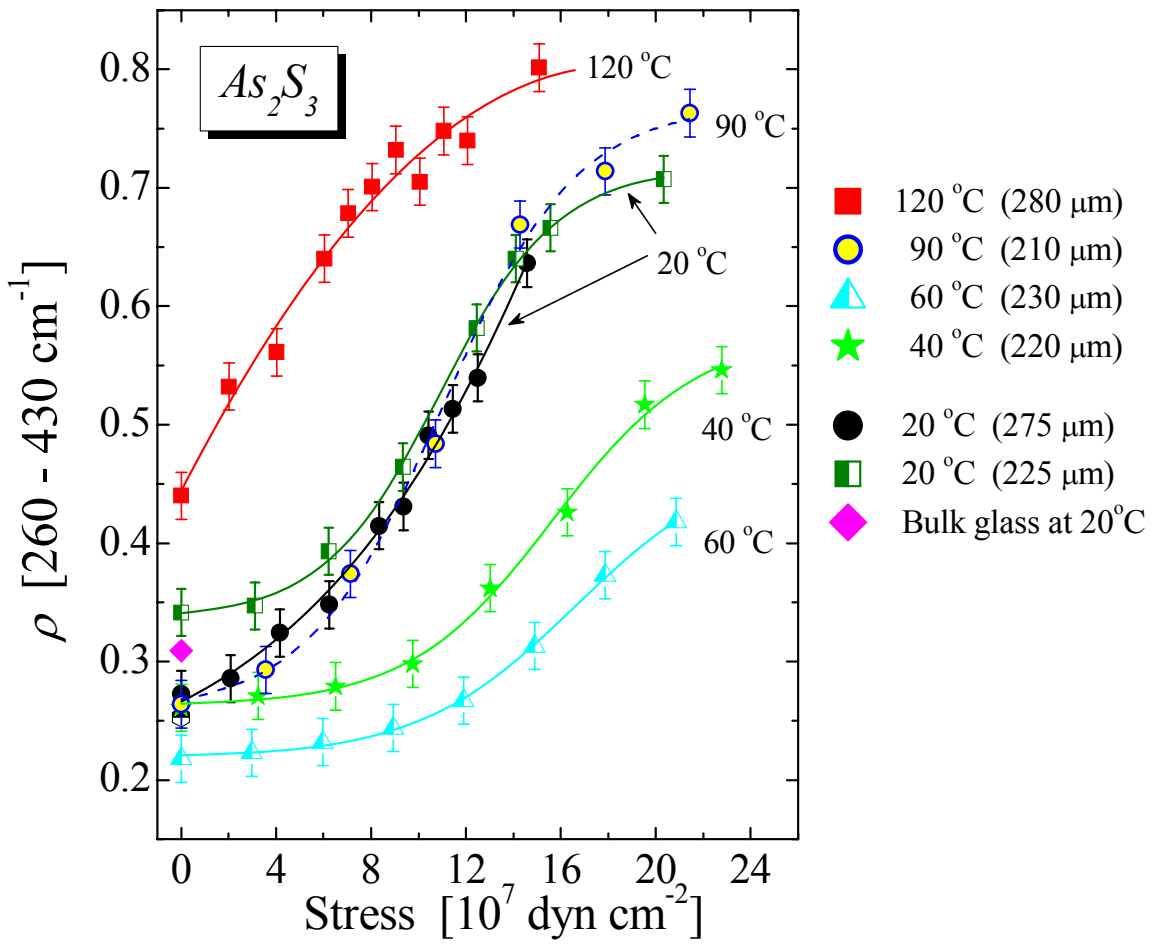

**Figure 3** Stress dependence of the depolarization ratio in the spectral region [260-430 cm$^{-1}$] for $As_2S_3$ fibers at various temperatures. Sigmoid dashed and solid lines are guides to the eye. The data illustrate the anomalous characteristics of the photoviscous effect through the reversal of the temperature dependence above 60 °C [40].

To quantify the reversal in the temperature dependence of the photoplastic effect we show in Fig. 5 the constant-stress (at $S$=10×10$^7$ dyn cm$^{-2}$) value of $\rho$ for various temperatures. The magnitude of $\rho$ is considered proportional to the photoplastic deformation of the fiber. For reasons to be discussed below the mechanism of photoinduced fiber elongation seem to be severely less effective near 60 °C.

The mechanisms involved in the plastic deformation of the $As_2S_3$ glass have been reported in detail elsewhere [40, 41] and will only briefly discussed here. Models where *intramolecular* bond rupture occurs have been proposed by Fritzsche [34(a)] (self-trapped exciton) and Yannopoulos [40] (opening and incorporation of $As_4S_4$ cagelike fragments into the glass structure). It was shown that in the framework of the second model [40] a quantitative estimation of the fiber elongation was provided. An *intermolecular* structural model (buckling model) [43(a)] has been mentioned as one possible candidate for intermolecular structural changes [33, 36, 39]. This model is based on the widely adopted idea that particular chalcogenides, including $As_2S_3$ glass, retain to some degree the layered structure of the crystalline phase. The inter-layer interactions (van der Waals bonds) are weak enough and hence are easily responsive to incident light. In all the aforementioned models, the specific atomic transformations involved engender structural anisotropy as is clearly evidenced in the depolarization ratio increase.

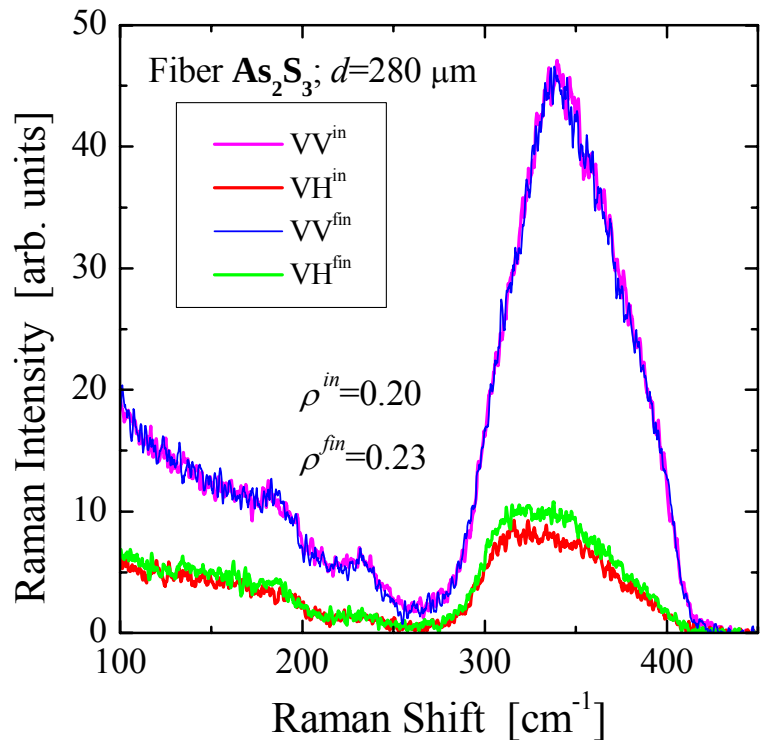

**Figure 4** VV and HV Stokes-side Raman spectra of an $As_2S_3$ fiber. The initial state ("in") corresponds to the spectrum at S=0. The final state ("fin") corresponds to the spectrum recorded after stressing (S=8×10$^7$ dyn cm$^{-1}$) the fiber in dark at 120 °C for 3 h and cooling back to ambient temperature.

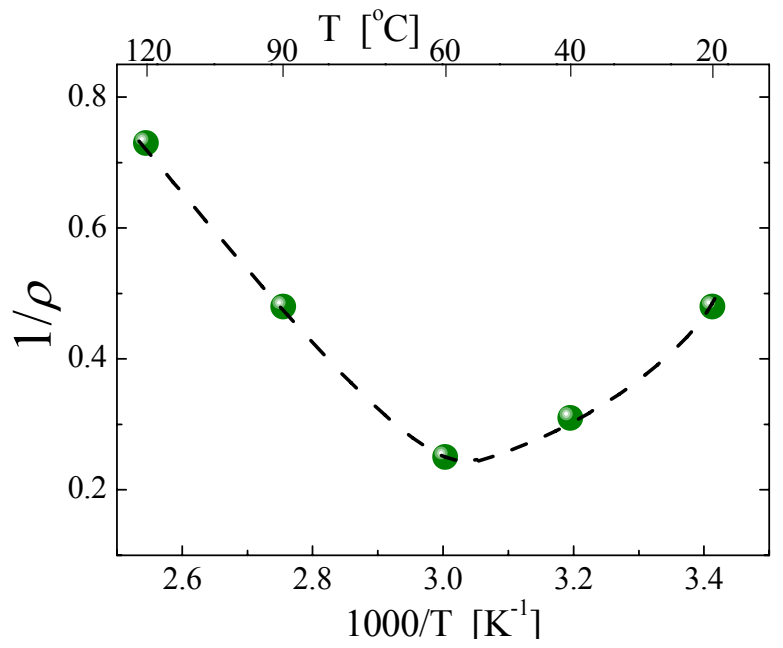

**Figure 5** The reversal of the photoplastic effect: temperature dependence of the depolarization ratio at constant stress.

More intriguing is the explanation of the temperature dependence of photoplastic changes shown in fig. 5. A simplistic idea to account for the decreased photoplastic changes up to 60 °C was based on the reduced geminate recombination rate due to increased mobility of photo-excited carriers [33(a), (b)]. For reasons explained elsewhere [41] the dramatic decrease of photo-induced viscosity cannot be accounted for, solely, by a decrease of the geminate recombination rate. The role of the process of scission and polymerization of $As_4S_4$ molecules must be significant [40]. To account for the reversal of photoplastic effects at T > 60 °C it would be instructive to recall structural studies of As-S glasses at high temperatures [43(b)]. This study provided experimental evidence of an increasing layer ordering with temperature. Therefore, the non–monotonic dependence on temperature of the $\rho$ vs. $S$ curves shown in Fig. 3 was considered to arise from the *competition* of two opposing effects [40]. The first one is the decreasing geminate recombination rate and the gradual elimination of $As_4S_4$ molecules discussed above. The second effect, namely ordering effects through enhanced lay-ering, has the opposite temperature trend compared to the previous mechanism, tending to facilitate the plastic deformation imposed by the external stress. The reversal observed in Fig. 5 at ~60 °C signifies the predominance of the second mechanism above this temperature. The enhancement of layer ordering is presumably the origin of the increased *S*=0 value of $\rho$ at 120 °C, (Fig. 3).

**3.3 Suppression of photoplastic effects in non-stoichiometric AsxS100-x glasses** To examine the generality of the photoplastic effect studied above for $As_2S_3$ glass, investigations were performed for glasses of more flexible structures containing the chainlike and ringlike fragments characteristic of sulfur rich mixtures. The $As_{25}S_{75}$ glass composition is a convenient one since its structure, being nanophase scale separated, contains appreciable fractions of $AsS_3$ pyramids, $S_8$ rings and $S_n$ chains [44]. Further, its high enough glass transition temperature ($T_g \approx 140$ °C) ensures stability at room temperature.

The same analysis of Raman spectra, as previously discussed, was followed. The obtained results for ambient temperature are reported in Fig. 1 where for comparison the stoichiometric glass data are shown. The $As_{25}S_{75}$ glass has a wider band-gap ($E_g$ = 2.55 eV) than the $As_2S_3$ glass, and therefore both the 647.1 nm (open circles) and the 514.5 nm (diamonds) laser lines were used as a sub-bandgap illuminating sources. Apparently, a drastic suppression of photoplastic effects is observed for the 647.1 nm laser energy in comparison with the $As_2S_3$ data. It is also obvious the ineffectiveness of the 514.5 nm laser line, which lies below but very close to the band-gap of the glass, to induce photoplastic effects.

The observed behavior of the non-stoichiometric glass was discussed by addressing the role of three possible factors involved: (i) the difference in their structure, (ii) the relation between incident light energy and band-gap energy, and (iii) the role of the glass transition temperature. The last two were discussed in detail in Refs. [40, 41]. However, the impact of atomic structure is now more obvious after examining the photoinduced structural changes of $As_xS_{100-x}$ glasses using resonant and off-resonant Raman scattering [44]. In brief, it was found [44] that near-bandgap light induces scission of $S_8$ rings and polymerization of the diradicals formed to $S_n$ chains. These atomic transformations from zero-dimensional units to sub-structures of higher dimensionality lead to a stiffer glass structure that exhibits reduced photoplastic ability. Finally, the lack of As-As bonds in S-rich glasses adds to the observed lack of photoplasticity of such glasses.

Concluding, low dimensional and highly constrained 3D network structures (with compact structures, lacking weak van der Waals bonds) are not expected to exhibit significant photoplastic effects. As we have already stressed [39, 41] the particular structural characteristics and the softness of *locally layered materials* maximize the possibility of structural changes that lead to the photoplastic deformation of the material.

**4 Effect of light on the elastic-plastic characteristics and rheological properties of amorphous chalcogenides: Photoinduced change of elastic modulus and nanohardness** In this section we will briefly outline experimental work related to photoplastic effects undertaken by *in situ* micro- and nanoindentation experiments. Microhardness was measured the microindentation method, which employs a diamond Vickers pyramid as an indenter. Significant photoplasticity of thin As-S(Se) films was revealed through microindentation kinetic experiments. It was found that the viscosity $\eta$ decreases from $\eta \geq 10^{16}$ Poise in the dark to $10^{12}$-$10^{13}$ Poise under illumination both for glasses of As-S and As-Se systems [26]. This magnitude of viscosity corresponds to that of the "dark" glass when heated close to $T_g$. Similar results were obtained for chalcogenide films of other compositions, i.e. for Sb-Se systems, that led us to suggest a common microscopic origin of the observed phenomenon. However, the ability of microindentation techniques to study photoinduced elastic-to-plastic transformations in glasses is limited by difficulties associated with monitoring the process of indenter penetration, the relatively large area of the indented surface, and the inability to calculate other elastic-plastic parameters but microhardness.

Depth-sensing instrumental indentation (nanoindentation) has become a common technique for measuring mechanical behavior of a large variety of materials, including their photomechanical response under light irradiation both for crystalline [45] and non-crystalline materials [46]. This method allows the recording of the penetration depth of a sharp indenter as a function of applied load. As a result the material hardness and its changes are directly determined by analysis of the corresponding load–displacement curves independently from the residual imprint. Static mechanical properties such as Young's modulus and local (nano) hardness can also be obtained from the unloading curve. An example of a nanoindentation experiment is shown in Fig. 6(a) for a 3 μm thick, as-deposited a-$As_2S_3$ film which was indented in dark, under illumination and after being illuminated at 532 nm (350 mW cm-2). Illumination times range from 1 to 50 min. Qualitatively similar results were found for an annealed $As_2S_3$ film. However, in the study of the annealed film the penetration depth of the indenter under illumination is much lower (almost by a factor of 2) than the corresponding one in the as-deposited film. This is a reasonable finding in view of the intramolecular model proposed to account for photoplastic changes [40]. Indeed, annealing reduces to a large extent the cagelike molecular units existing in the as-deposited film structure, hence decreasing the density of As-As homoatomic bonds which are considered responsible for the photoplastic effect [40]. Similar types of experiments were performed in dark for a metal (Pb), an organic polymer (polycarbonate) and amorphous $SiO_2$; the results are shown in Fig. 6(b). Figure 6(c).

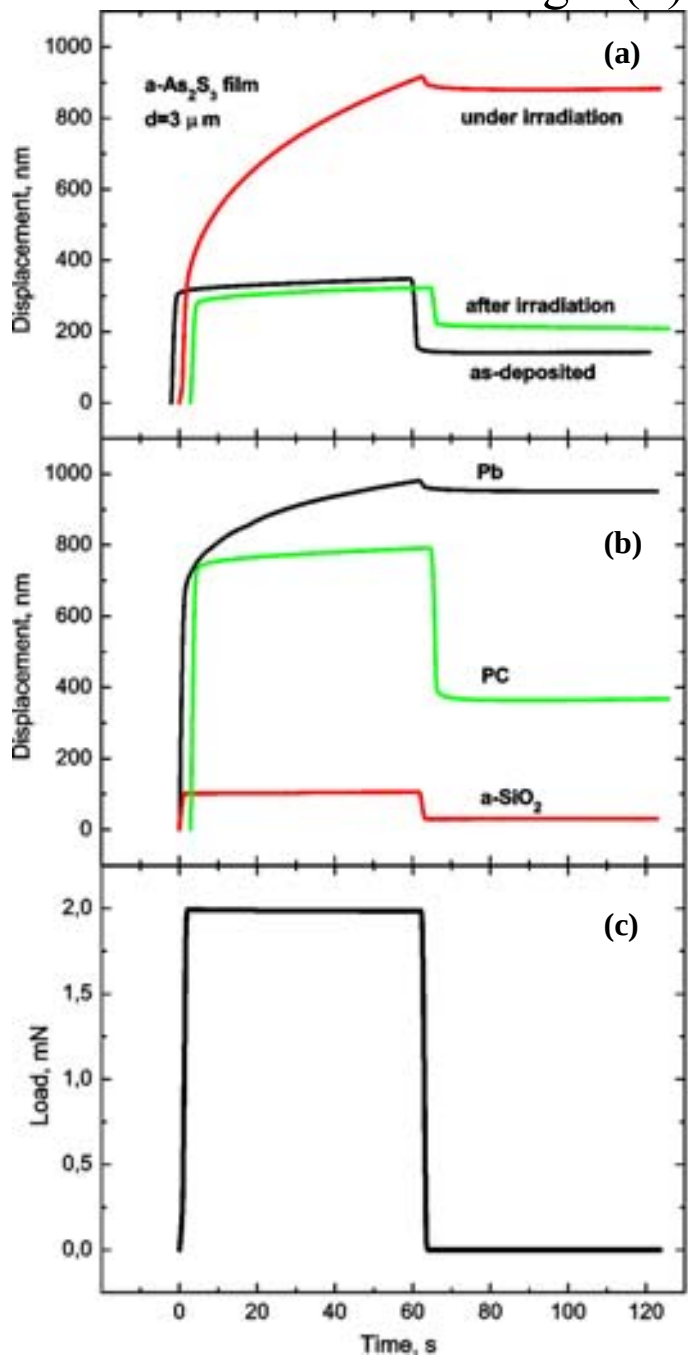

**Figure 6** Nanoidentation curves for (a) $As_2S_3$, (b) Pb, PC, $SiO_2$, (c) loading scheme. For clarity, the abscissa of some curves is shifted slightly. See text for details.

presents the schematic diagram illustrating the variation in the applied load in the course of nanoindentation. All materials shown in Fig. 6 exhibit similar behavior upon indentation. The relevant parameters determined experimentally for the $As_2S_3$ film are tabulated in Table I in Ref. [47]. Under bandgap light irradiation the indentation depth vs. time curve for $As_2S_3$ film is different compared with the curves of the dark and post-illuminated films. After a first sudden increase (exhibited in all cases) the indentation depth keeps on increasing albeit with a time dependent rate. This finding indicates decrease in hardness and viscosity and increase in plasticity. The striking similarity between the indentation depth vs. time curve for illuminated $As_2S_3$ and that of bulk Pb, the most representative example of a plastic material, shows that illumination induces a transition from a brittle to a plastic state in the structure of a-$As_2S_3$.

Surprisingly, the Young modulus of the film under irradiation is higher than that of the virgin state and decreases only slightly (~10%) for annealed film showing that the material becomes stiffer under light. While both the as-deposited and annealed films show a drastic decrease of nanohardness, upon illumination, the corresponding change of the elastic modulus is quite small but measurable. This unexpected finding has its origin to the difference in the structure between the as-deposited and the annealed films. The latter, undergoes irreversible structural modifications due to the scission and incorporation of cagelike units into the network glass structure. Annealing, brings the structure of the as-deposited film closer to that of the corresponding bulk glass. As it becomes evident from Fig. 6, the region of a purely elastic response persists under illumination even in the absence of loading. On the other hand, the viscoelastic depth recovery was found similar in dark and after irradiation. This finding shows that the film undergoes transformation to a plastic but not a fluid state. The same behavior was obtained for annealed films.

## 5 Related phenomena stimulated by the photoplastic effect

### 5.1 Internal stress generation and relaxation

Manifestations of the photoplastic effect are evident through photorelaxation of internal stress. It is well known that thin films condensed onto substrates posses built-in internal stresses which appear during the deposition procedure as a result of the different thermal expansion coefficients of the film and the substrate, structural instability, as well as other reasons. It has been shown that internal stresses disappear when the material is exposed by bandgap light [26(a)]. Let us consider the typical case of As-S films deposited on glass substrates by thermal evaporation (Fig. 7(a)) which shows that while as-deposited As-S films are characterized by compressive stresses (line 1), film annealing results in the emergence of tensile stresses (line 2). Bandgap illumination (50-300 mW $cm^{-2}$) decreases stress magnitude to zero, which subsequently increases slightly in dark. On repeating laser on-off cycles, reversible changes in the internal stress from this level to zero takes place (line 3).

The particularly interesting feature of the above results is that the magnitude of the internal stress in films irradiated to full saturation is very close to zero. This finding depends neither on thermal history and preparation method nor on the temperature during the process of irradiation (Fig. 7(b)). Analysis shows that photosoftening is at the origin of this effect. Indeed, stress relaxation can be satisfactorily described by the Maxwell relation $\sigma = \sigma_0 \cdot \exp(-t/\tau)$, with a relaxation time $\tau$=100-1000s depending on the illumination intensity. This is a signature of the viscous flow of Newtonian liquid with a viscosity of $\eta \sim 10^{13}$ Poise. The result agrees with the glass viscosity determined from nanoindentation experiments described in section 4. The internal stress relaxation to zero upon bandgap illumination described above was achieved by unpolarized light (scalar effect). Further studies revealed that illumination of an originally isotropic, internal-stress-free, amorphous chalcogenide film with polarized light causes generation (growth) of stresses (vectoral effect) [17].

**5.2 Polarization-dependent anisotropic mass-transport phenomena and surface photodeformations** As mentioned in Section 2.7 formation of giant relief modulations in a-$As_2S_3$ has been reported upon intense bandgap light illumination [30(a)] attributed to the generation of polarizable defects (due to photoinduced softening or photoviscous effect) which drift under the optical field gradient force forming surface relief modulations. In more

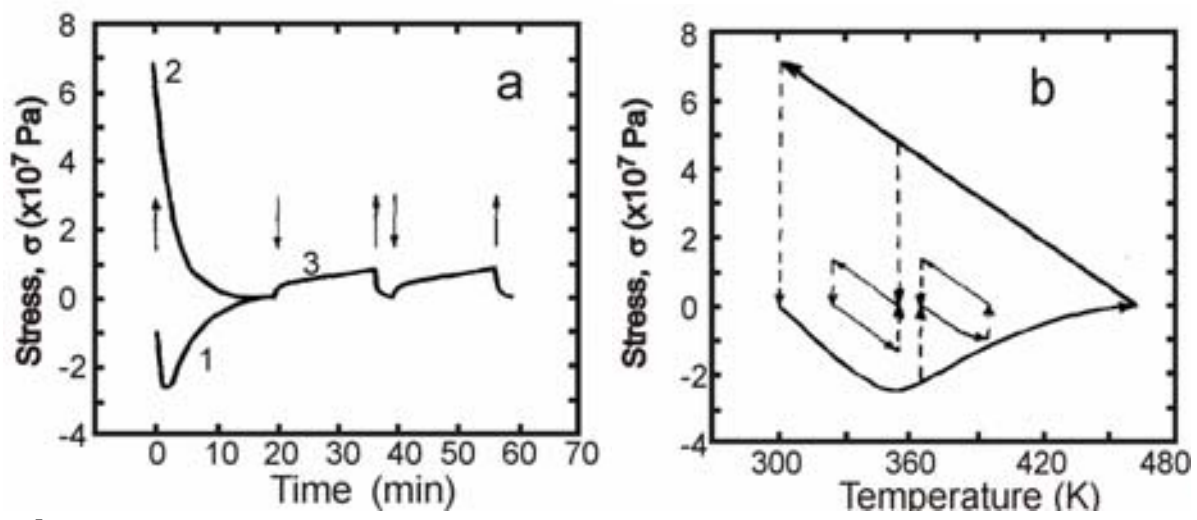

**Figure 7** (a) Photoinduced stress relaxation in $As_2S_3$; the arrows show the beginning and the end of exposure. (b) Dependence of the induced stress on temperature variation (solid lines) and irradiation (dashed lines) [26(a),(c)].

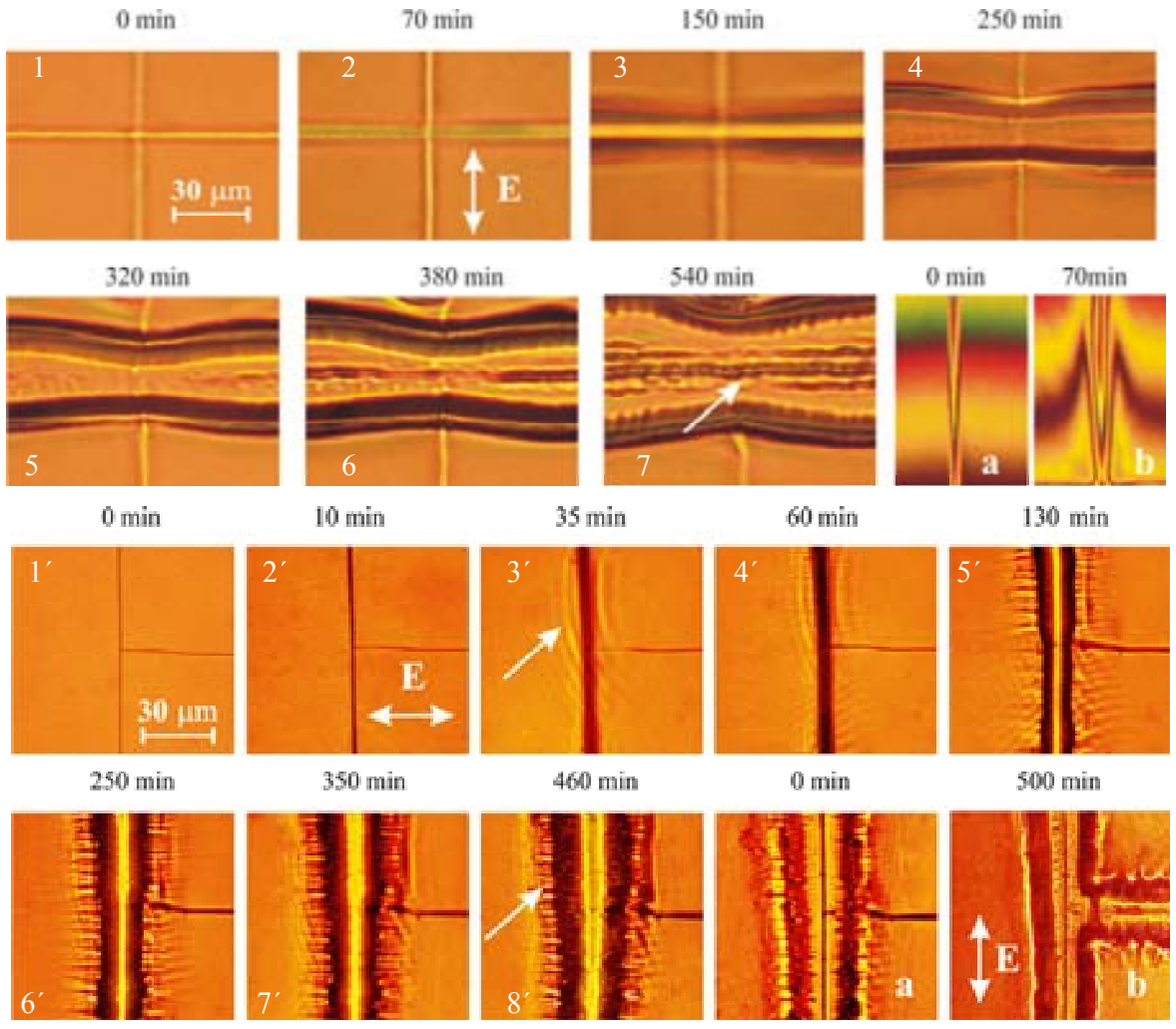

**Figure 8** *Upper part.* Reflected light microphotographs showing anisotropic photoinduced widening of a channel in as-deposited 2 $\mu$m thick $As_{20}Se_{80}$ film. Images (a) and (b) shows cross-sections (interferograms) of the channel before and after irradiation for 70 min, respectively. *Lower part.* Same as above for the crack on the surface of an annealed 3 $\mu$m thick $As_2S_3$ film. Images (a) and (b) were obtained after rotation of the direction of polarization by 90° showing initial and final (500 min of irradiation) channel morphologies. The double arrow denotes the direction of the polarization.

recent studies, giant, laser induced, polarization-sensitive mass-transport was found to take place under irradiation at moderate light intensity and long exposure times [31, 32]. As illustrated in Fig. 8 (upper part), when two channels perpendicularly scratched on an $As_{20}Se_{80}$ film are illuminated by linearly polarized light (633 nm, 250 mW·cm-2) the width of the channel whose orientation is orthogonal to the polarization direction of the incident light increases significantly, almost by a factor of 30. A cross-section analysis of the widened channel shows that the initial V-shaped profile of this channel (Fig. 8(a)) changes after illumination to an M-shaped one (Fig. 8(b)). The channel expansion process is accompanied by the formation of giant pill-ups at its edges; appreciable mass transport of the material occurs. On the contrary, the width of the channel oriented along the polarization direction of the electric field of the incident light remains almost unchanged.

It should be emphasized that this anisotropic effect described above does not depend on channel's depth; it suffices to be even a crack. Figure 8 (lower part) shows the effect of the photodependence of crack growth on an annealed $As_2S_3$ which again reveals the anisotropic response to illumination of linearly polarized light (532 nm, 350 mW·cm$^{-2}$) [47]. The width of the channel oriented parallel to the direction of the electric field of the incident light increases only when the latter is rotated by 90° (see Fig. 8(b), lower part).

It should be noted that that illumination of a chalcogenide surfaced by circularly polarized light results in the expansion of both perpendicularly arranged channels while a heating the film to near $T_g$ would lead to the healing of both channels The same behavior was detected in bulk glasses of $As_2S_3$ and $As_{20}Se_{80}$ [32]. Therefore, the photoinduced effect observed using linearly polarized light enables us to exclude a thermal origin of the effect. Studying the compositional dependence of the aforementioned effect it is revealed the chalcogenide compositions that exhibit the most appreciable changes are $As_{20}Se_{80}$ and $As_{40}S_{60}$ in both the bulk and the thin film forms. The criterion for monitoring the magnitude of the effect was the increase rate of the channel width.

### 5.3 Spontaneous surface relief formation

Macro-scale anisotropic deformations on the surface of As-Se films has been observed by bandgap irradiation of linearly-polarized light [48]. Unfocused, relatively homogeneous, low-intensity (20 mW cm$^{-2}$) illumination of Se-rich films results in an unusual spontaneous relief formation, superimposed to the original structure of the surface. Under long-time exposure the spatial scale of these relief deformations acquires macroscopic size and can be observed even under an optical microscope. The deformation is polarization sensitive and changes shape by rotating the electric field polarization plane.

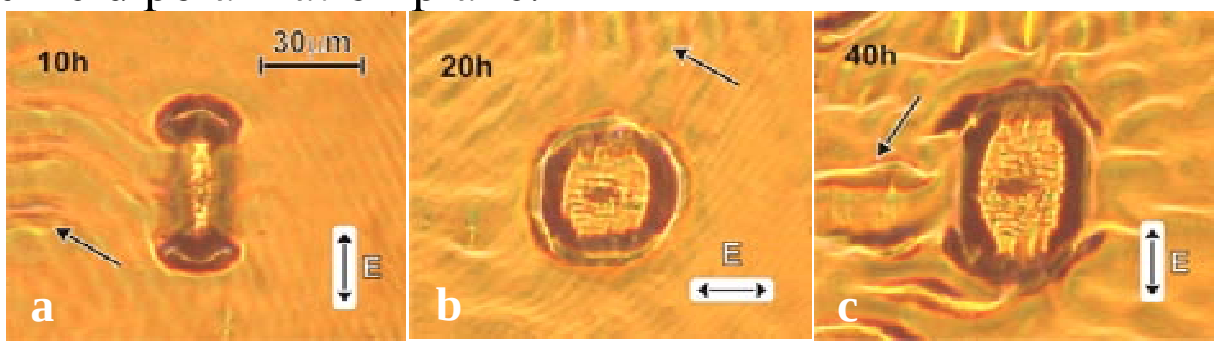

**Figure 9** Reflected light microphotographs showing the evolution of the morphological changes on the surface of an as-deposited $As_{20}Se_{80}$ film under irradiation by a linearly polarized light (633 nm, 20 mW cm$^{-2}$, laser spot ~1.5 mm). The double arrow denotes the direction of the polarization [48].

Figure 9 illustrates optical microscope images of evolution of surface morphology changes under illumination for an a-$As_{20}Se_{80}$ film (which experiences the highest extent of such changes). The giant relief deformations shown here were obtained after periodic rotation of the sample by 90° in respect to the direction of the polarization vector of the incident light. Three well-distinguished morphologies of the film surface are seen in Fig. 9. At the early stages of illumination a limited surface relief appears, see Fig. 9(a). The orientation of the relief "strips" is always orthogonal as shown by black arrows in Fig. 9 to the incident light polarization vector shown by white arrow.

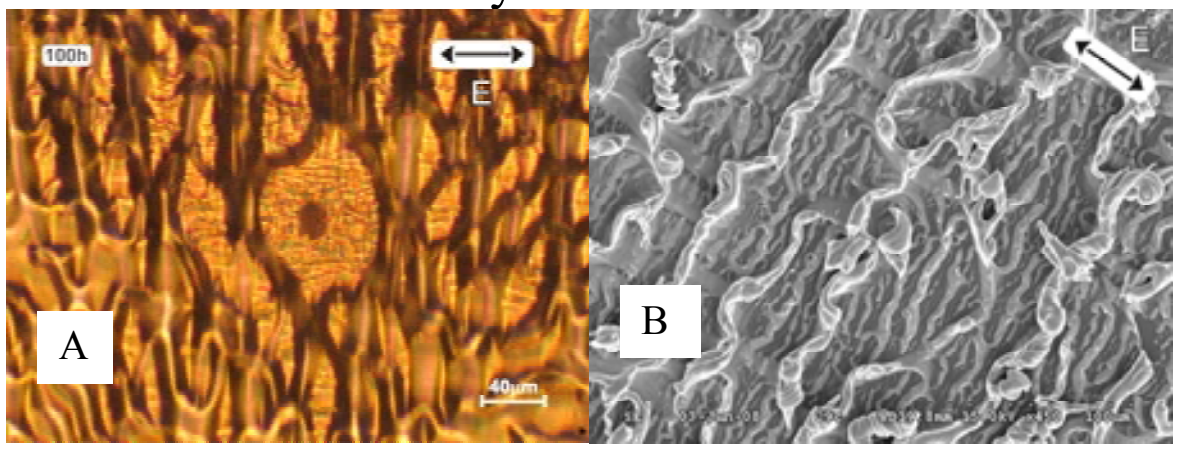

**Figure 10** (a) Reflected light microphotographs showing giant photoinduced anisotropic deformation of the surface of $As_{20}Se_{80}$ film after long-term irradiation by linearly polarized light (633 nm, 20 mW cm$^{-2}$, laser spot ~1.5 mm). (b) SEM image showing surface relief deformations and flakes with twisted (spiral) morphology being oriented orthogonally to the direction of the electric field of light. The double arrow denotes the direction of the polarization.

Another feature of the surface morphology evolution is a pit formation which is growing along the light polarization. Further, a hole inside the center of the formed pit is evolving which is continuously increasing under illumination (see Fig. 9 b, c). For longer illumination times the pit size and the surface relief increase continuously. After 100 h of illumination the height of the latter exceeds one half of the film thickness, see Fig. 10(a). Impressive changes on film's surface were detected after prolonged (more than 200 hours) illumination. The film is completely damaged and the formed thin flakes experience anisotropic deformation. In particular, flakes acquire twisted (spiral) morphology and orient orthogonally to the direction of the electric field of light as illustrated in the scanning electron microscope image in Fig. 10(b). There might be some relation between this shape effect and the directed bending effect reported for azobenzene materials [9(b), (c)] and more recently for $As_2S_3$ sub-mm sized flakes [33(c)].

**6 Conspectus** In the preceding brief review we have tried to outline some characteristic photoinduced effects concerning mechanical, rheological, and elastic properties of amorphous chalcogenides. Epigrammatic reference was given to similar effects exhibited by azobenzene-containing materials. Although attempt was made to cover in a representative way a wide range of such phenomena there are obviously many more articles not cited here due to the limited space of the present review. Critically assessing these effects one can reach the conclusion that most of the effects discussed in Section 2 seem to have a common origin, i.e. is the athermal increase of viscosity or equivalently the emergence of viscoeleastic behavior under light illumination. This can alternatively be viewed as a brittle-to-plastic transition for the illuminated material.

Some representative results on photoplastic effects of the authors' work were presented. Raman studies dealing with the pursuit of the microscopic origin of the structural changes can offer valuable information on both the intra- and intermolecular levels of structural organization. In particular, it was shown that the Raman results are compatible with self-trapped exciton models, layer unfolding processes, and scission of homopolar As-As bonds. Micro- and nanoidenttaion studies under in situ illumination provide also a direct view of the changes elastic-plastic parameters lending solid evidence on the viscoelastic nature of photoplastic effects.

Concluding, morphic effects, i.e. changes in the shape and/or the surface morphology frequently appear under illumination of linearly polarized light either in the presence or absence of external mechanical stimuli. Light intensity, polarization and illumination time have distinct roles in photoplastic effects. Photoprocessed structures are envisaged to be of high technological importance for micro- or nano-manipulation. The transformation of photoplastic effects into commercially viable applications will be decided from the improvement of our knowledge on the basic mechanisms underlying the effect. However, even if this still seems remote, these photo-induced plastic effects deserve further study since they are of fundamental interest in their own right.

**Acknowledgement** SNY wishes to thank G. N. Papatheodorou and D. Kastrissios who contributed to the work presented in Section 3 of the paper. Thanks are also due to H. Fritzsche for bringing to our attention the photoplastic effects in fibers. SNY also acknowledges financial support of the "PENED-03/EΔ-887" project which is co-funded: 75% from the European Union and 25% of public financing from the Greek State, Ministry of Development–GSRT and the Hellenic Telecommunications Organization (OTE S.A.). MLT appreciates a great contribution of his collaborator A. G. Anchugin (who passed away in his early years) to the initial stage of joint research of photoplastic effect in amorphous chalcogenide films. The nanoindentation data were obtained in collaboration with Dr. S. N. Dub of the Bakul Institute for Superhard Materials, National Academy of Sciences of Ukraine. MLT acknowledges financial support from the International Visegrad Fund (Contract No 50810548).